# Calorimetry of a Bose-Einstein condensed photon gas

Tobias Damm,  Julian Schmitt,  Qi Liang¶, David Dung, Frank Vewinger, Martin Weitz
& Jan Klaers*§

*Institut für Angewandte Physik, Universität Bonn, Wegelerstr. 8, 53115 Bonn, Germany*

¶present address: *Atominstitut, TU Wien, Stadionallee 2, 1020 Vienna, Austria*

*present address: *Institute of Quantum Electronics, ETH Zürich, Auguste-Piccard-Hof 1, 8093 Zürich, Switzerland*

§corresponding author:  jklaers@phys.ethz.ch

**Phase transitions, as the condensation of a gas to a liquid, are often revealed by a discontinuous behavior of thermodynamic quantities. For liquid Helium, for example, a divergence of the specific heat signals the transition from the normal fluid to the superfluid state. Apart from liquid helium, determining the specific heat of a Bose gas has proven to be a challenging task, for example for ultracold atomic Bose gases. Here we examine the thermodynamic behavior of a trapped two-dimensional photon gas, a system that allows us to spectroscopically determine the specific heat and the entropy of a nearly ideal Bose gas from the classical high temperature to the Bose-condensed quantum regime. The critical behavior at the phase transition is clearly revealed by a cusp singularity of the specific heat. Regarded as a test of quantum statistical mechanics, our results demonstrate a quantitative agreement with its predictions at the microscopic level.**

Below 2.2K liquid helium shows peculiar hydrodynamic properties, like a flow without viscosity, the fountain effect or the formation of vortices[1]. This transition from a normal fluid to a superfluid has been named λ-transition, which originates from the fact that plotting the heat capacity versus temperature[2] results in a graph resembling the Greek letter λ. Soon after this discovery, it has been proposed that superfluid helium forms a macroscopic matter wave as a consequence of Bose-Einstein condensation[3], which describes the condensation of the ideal (interaction-free) Bose gas at low temperatures due to quantum statistics[4]. This idea proved to be fruitful despite the fact that liquid helium is far from a system of interaction-free particles[5]. The impressive progress in the cooling of dilute atomic gases has paved the way to realize weakly-interacting Bose gases at nano-Kelvin temperatures[6,7]. Here, the relation to Bose-Einstein condensation has been immediately clear. Interestingly, in contrast to liquid helium and a recent measurement of a strongly interacting atomic Fermi gas[8] these systems have not allowed for detailed calorimetric studies up to now[9]. Evidence for a non-classical specific heat has been reported[10,11], but the accuracy obtained in experiments with weakly interacting atomic Bose gases has not been sufficient for an unambiguous determination of the temperature dependence of the heat capacity.

Following ultracold atomic Bose gases, other physical systems have been demonstrated to undergo Bose-Einstein condensation, for example gases of exciton-polaritons[12,13], magnons[14] and, in previous work of our group, photons[15-17]. In contrast to a three-dimensional thermal photon gas as Planck's blackbody radiation, photons can exhibit Bose-Einstein condensation, if the thermalization process is restricted to two motional degrees of freedom. Experimentally, this situation has been realized in a microcavity enclosing a dye medium, designated as a room temperature heat bath for the photon gas. Detailed experimental studies of the thermalization[16] and condensation process[15], as well as the quantum statistics of the photon condensate[17], have revealed the signatures of an almost ideal Bose gas.

Here, we report a measurement of the calorimetric properties of a Bose-Einstein condensed photon gas, in particular, the temperature dependence of the specific heat and entropy from the classical high temperature to the quantum degenerate regime at low temperatures. At the phase transition, the observed specific heat shows a cusp singularity, illustrating critical behavior for a photon gas analogous to the λ-transition of liquid helium.

**Results - Two-dimensional photon gas in a dye-microcavity** In our experiment (see Fig. 1a), photons are captured inside a microcavity consisting of two spherically curved mirrors while repeatedly being absorbed and re-emitted by the embedded dye medium. The cavity length is of the same order as the wavelength itself, which causes a large frequency gap between the longitudinal resonator modes (free spectral range), comparable to the emission bandwidth of the dye molecules, see Fig. 1b. In this situation, the resonator becomes populated by photons of a single longitudinal mode number $q$ only, for example $q=8$. While the longitudinal mode number is frozen out, the photons may populate a multitude of transversally excited cavity modes, e.g. the $TEM_{8xy}$ sub-spectrum, which makes the photon gas effectively two-dimensional. The photon energy-momentum relation acquires a quadratic form, resembling that of a massive particle, and a trapping potential for the photon gas is induced by the mirror curvature. One can show that the photon gas confined in the resonator is formally equivalent to a harmonically trapped two-dimensional gas of massive bosons[18,19], described by the dispersion

$$E \approx m(c/n)^2 + (\hbar^2/2m)(k_x^2 + k_y^2) + (m\Omega^2/2)(x^2 + y^2) \,, \qquad (1)$$

with spatial coordinates $x,y$, transverse wavevector components $k_x,k_y$, trapping frequency $\Omega$ and an effective mass $m=\hbar\omega_c(n/c)^2$, where $\hbar\omega_c$ is the photon energy in the cavity ground mode with $n$ as the refractive index of the medium and $c$ as the vacuum speed of light. Thermal equilibrium of the photon gas with the cavity environment at room temperature is achieved via repeated absorption and emission processes by the dye molecules, which establishes a thermal contact between photon gas and optical medium[18,20,21]. Other than in a blackbody radiator, the thermalization process allows for an independent adjustment of temperature and photon number, for example by (initial) optical pumping, which eventually goes back to a separation of energy scales of photon energy and thermal energy. In our experiment, the Bose-Einstein condensation is triggered by increasing the photon number above the saturation level at a given temperature. The corresponding critical particle number is given by[22-24]

$$N_c \approx (\pi^2/3)(k_B T/\hbar\Omega)^2 \,, \qquad (2)$$

which for typical experimental parameters corresponds to $N_c \approx 90{,}000$.

One of the benefits of the given experimental system is that one can easily interpolate between equilibrium and non-equilibrium experimental conditions. Parameters, such as mirror reflectivity, dye concentration, cavity tuning and pump geometry, can be chosen such that gain and dissipation either significantly contribute to the system dynamics or effectively drop out of it[16,20,25-27]. In this study, we have concentrated on the equilibrium properties of the system. Details on the corresponding experimental parameters can be found in the Methods section. In our experiments, the optical medium is pumped with a spectrally off-resonant laser source at a wavelength of $\lambda_{exc}$=532 nm, having a relatively large beam diameter of approximately 150μm to keep the excitation of the medium nearly spatially homogeneous. Two acousto-optical modulators (AOM) are used to chop the pump light to long pulses of 400 ns length with a repetition rate of 400 Hz. The AOMs further control the intensity of each light pulse which allows us to adjust the average photon number with respect to the critical photon number. In our experiment, we quickly ramp the total photon number from typically 30,000 to 550,000 photons within 250 ms.

The thermodynamic properties of the system are experimentally accessible by spectroscopic means. To obtain the intra-cavity spectral distribution of the photons we measure the spectral distribution outside the resonator and divide by the (wavelength-dependent) transmission coefficient of the mirrors. For this, the emitted cavity light is collimated and coupled to a monochromator (4f-setup) where it is spectrally decomposed by a diffraction grating (1200 grooves/mm) giving an overall resolution of 0.5 nm. With this setup, it is possible to capture the photon gas spectrum in the wavelength region starting from ≈550 nm to the cavity cut-off at $\lambda_c$ =580 nm corresponding to an energy range of ≈4 $k_B T$. The latter comprises the condensate population and ≈95% of photons in excited cavity modes. The ≈5% most energetic photons of the thermal cloud are not experimentally resolvable with the present setup.

Figure 2a shows spectra obtained at fixed temperature $T$=300K for total photon numbers ranging from $N$≈30,000 to $N$≈550,000 (varying chemical potential). The observed spectra generally are in good agreement with Bose-Einstein distributions, with residual deviations at the lower photon energy part due to imperfect reabsorption. The measured energy distributions can be used to obtain full thermodynamic information of the two-dimensional photon gas. All of the derived quantities will be measured for constant volume, which here means constant (inverse squared) trapping frequency, and fixed absolute temperature $T$. As a first step, we have determined the condensate fraction $n_0/N$ as a function of the reduced temperature $T/T_c$, see Fig. 2b. From eq. (2) follows that the reduced temperature $T/T_c$ is related to the total particle number by $T/T_c=(N_c/N)^{1/2}$, with the total particle number $N$ being obtained by integrating over the spectrum. The experimentally derived condensate fractions are slightly below the theoretical expectations shown by the solid line in Fig. 2b, describing an inverse parabola $n_0/N=1-(T/T_c)^2$. This stems from an imperfect saturation of the population in excited photon modes which has already been observed in previous measurements[15] and potentially originates from a weak (thermo-optically induced) photon self-interaction[28].

**Caloric and entropic properties** We next determine the average energy per photon $U/N$, with the zero point of the energy scale being set to the energy $\hbar\omega_c = hc/\lambda_c$ of the cavity ground state (TEM$_{00}$ mode), corresponding to a condensate wavelength of $\lambda_c=580$nm. Based on the experimentally obtained spectral photon distribution $n(\lambda)$ in the wavelength regime $\lambda \approx 550\text{-}580$ nm, we extrapolate the total internal energy to be $U \approx \kappa \times \int_{550\text{nm}}^{580\text{nm}} n(\lambda)\, hc\, (\lambda^{-1}-\lambda_c^{-1})\, d\lambda$, with $h$ as Planck constant and $c$ as the vacuum speed of light. The extrapolation factor $\kappa$ is uniquely determined by the assumption that the spectral distribution continues to be Boltzmann-like in the experimentally not resolved wavelength regime $\lambda < 550$ nm, containing the $\approx 5\%$ most energetic photons of the thermal cloud. The latter sets a value of $\kappa=U/U(\lambda>550\text{nm})\approx 1.19$, with $U(\lambda>550\text{nm})$ denoting the energy contribution of photons with wavelength $\lambda>550$nm.

In Figure 3a, the average energy $U/N$ normalized to the characteristic energy at criticality $k_B T_c$ is plotted versus the reduced temperature $T/T_c$ showing good agreement with the theoretical expectations for an ideal Bose gas (solid lines). At higher temperatures $T>T_c$, the energy shows a linear scaling with temperature, as expected in the classical limit where Maxwell-Boltzmann statistics applies. In the vicinity of the condensation threshold, the energy curve changes slope, as is more clearly revealed in the graph of the heat capacity in Fig. 3b. The data points are obtained by numerically differentiating the measured energy curve with respect to temperature, following $C=\partial U/\partial T=k_B\, \partial(U/k_B T_c)/\partial(T/T_c)$. The obtained heat capacity data (circles) shows the characteristic $\lambda$-like shape as predicted theoretically (solid lines). In the high temperature (classical) regime, $C$ reaches a limiting value of $2k_B$ per photon. The latter stems from the four degrees of freedom, two kinetic ($k_x,k_y$) and two potential ($x,y$), that quadratically enter the photon energy of eq. (1) and each contribute with $k_B/2$ to the specific heat (equipartition theorem). At criticality, the heat capacity shows a cusp with a maximum value of $C(T_c)/N$=(3.8±0.3) $k_B$, slightly below the theoretically expected value in the thermodynamic limit of 6$\zeta$(3)/$\zeta$(2) $k_B$≈4.38 $k_B$ (Ref. 22-24). Most of this discrepancy can be explained by finite size effects. From an exact numerical evaluation of the Bose-Einstein distribution function, one can obtain the specific heat for finite system sizes (solid lines in Fig 3b). For the given photon numbers near $N$≈90,000 at criticality, finite size effects reduce the specific heat maximum to a value of 4.11 $k_B$, which agrees with the measured value within the experimental uncertainties. For $T<T_c$, the heat capacity monotonically decreases for smaller temperatures, dropping below the classical value of 2$k_B$ per photon at $T/T_c$≈0.7, and being consistent with reaching zero at $T$=0, as would be demanded by the third law of thermodynamics. Presently, the minimum achieved temperature is $T/T_c$≈0.4, corresponding to condensate fractions of up to 84%, due to limited available pump power.

The specific heat furthermore gives access to other thermodynamic quantities, as the entropy, which can be determined by the integral $S(T)=\int_0^T C(T)/T \, dT$, see Fig. 3c for corresponding data. As the specific heat $C(T)$ is not known for temperatures below $T/T_c$≈0.4, there is one free parameter, the constant offset $S(T/T_c$≈0.4) that needs to be set in order to evaluate the integral and to match the experimental data (circles) to the theoretical prediction (solid line). The entropy curve monotonically decreases with decreasing temperature, reaching a minimum value of ≈0.2$k_B$ per photon for the lowest obtained temperature. Although we presently cannot access temperatures closer to zero, the observed drop-off of the entropy curve towards lower temperatures is in accordance with the third law of thermodynamics. Note that the entropy per particle (as a function of chemical potential) also has been experimentally determined for a trapped two-dimensional atomic Bose gas, reaching entropies as low as 0.06(1)$k_B$ (Ref. 29).

**Discussion -** For a gas of non-interacting bosons, which in good approximation is realized in our experiment as demonstrated by the results given in Fig. 2 and 3, one can readily link the internal energy of the gas to its pressure. In general, in the presence of a trapping potential the pressure of a gas becomes position-dependent and thus cannot serve as a global thermodynamic variable. To account for this, it has been proposed to use two global conjugated variables, harmonic volume $V$ and harmonic pressure $P$ respectively[30,11]. For a harmonically trapped two-dimensional gas, the harmonic volume is defined as $V=\Omega^{-2}$, with $\Omega$ as the trapping frequency. This quantity does not have the physical units of a volume, however, it shares the same scaling with the trapping geometry as the "true" volume of the confined gas. The harmonic pressure then follows via the usual thermodynamic relation $P=-\partial \phi_G/\partial V$, where $\phi_G$ is the grand potential. For a two-dimensional trapped gas of non-interacting bosons, the pressure can be shown to be related to the internal energy via $P=(1/2)\Omega^2 U$. Thus, Fig. 3b here does not only describe the energy as a function of temperature $U=U(T/T_c)$ (caloric equation of state), but also delivers the pressure dependence $P=P(T/T_c)$ (thermal equation of state) for a given harmonic volume $V$ or trapping frequency $\Omega$.

To conclude, we have determined calorimetric properties of a Bose-Einstein condensed photon gas, in particular the temperature dependence of energy, heat capacity and entropy. Critical behavior of the photon gas is clearly demonstrated by a cusp in the specific heat curve at the condensation threshold. For the chosen experimental conditions, e.g. sufficient photon reabsorption and nearly homogenous pump geometry, we do not observe significant deviations from the theoretically expected behavior of a fully equilibrated Bose gas. In comparison to other systems exhibiting Bose-Einstein condensation, the here investigated photon gas comes closest to an ideal, i.e. interaction-free, gas of bosons, allowing to match experimental results with precise, and even exact, theoretical predictions. In particular, we find that the experimentally determined specific heat of the photon gas agrees with the quantum statistical predictions down to the level of finite size effects. For the future, the spectroscopic calorimetry of a quantum-degenerate photon gas could lead to new experimental schemes for precision measurements of thermodynamic quantities as the Boltzmann constant[31,32].

**Methods - Microcavity set-up**    In our experiment, photons are stored inside a microcavity build up from two gyro-quality mirrors. The dielectric, spherically shaped mirrors have a reflectivity in excess of $r \approx 0.99997$ in the relevant wavelength region of this experiment ($\lambda$=530-590 nm), providing a cavity finesse of order of $F \approx 10^5$ for the empty cavity. Both mirrors share the same radius of curvature of $R$=1 m and are typically separated by $D_0 \approx 1.7$ μm, corresponding to a free spectral range of order $\Delta\lambda_{FSR} \approx 100$ nm which is comparable to the spectral width of the dye emission. In this situation, the dye emission is restricted to cavity modes with longitudinal wave number $q$=8, effectively reducing the thermalization dynamics of the photon gas to the remaining two transversal mode number. The effective mass and trapping frequency introduced in equation (1) depend on the cavity geometry and typically take values of $m \approx 6.7 \cdot 10^{-36}$ kg and $\Omega \approx 2\pi \cdot 36.5$ GHz in our experiment. The cavity geometry is stabilized passively, by mechanical contact of the two mirrors which strongly damps fast mechanical oscillations, as well as actively by utilizing a piezo translation stage which counteracts long time drifts of the resonance.

As a heat bath for the photon gas, we use a filtered solution of $10^{-3}$ mol/l Rhodamine 6G in ethylene glycol (fluorescence quantum yield $\eta \approx 0.95$, index of refraction of the solvent $n = 1.43$). This dye fulfills the Kennard-Stepanov law $B_{21}(\omega)/B_{12}(\omega) \approx \exp(-(\omega-\omega_{ZPL})/k_B T)$, relating the Einstein coefficients of absorption $B_{21}(\omega)$ and emission $B_{21}(\omega)$ at a given frequency $\omega$ to the Boltzmann factor of that frequency ($\omega_{ZPL}$ is the zero-phonon line of the dye), which is essential for the light-matter thermalization process. The optical medium is spatially homogeneously pumped by a spectrally off-resonant laser system at a wavelength of $\lambda_{exc}$=532 nm under an angle of $\approx 45°$ with respect to the optical axis exploiting the first reflectivity minimum at higher angles of incidence. To avoid excess population of long-lived dye triplet states and photo bleaching two acousto-optical modulators (AOM) are used to chop the pump light to pulses of 400 ns length with a repetition rate of 400 Hz.

**Acknowledgements**

We thank James Anglin and Henk Stoof for valuable discussions. This work has been financially supported by the DFG (We1748-17) and the ERC (INPEC).

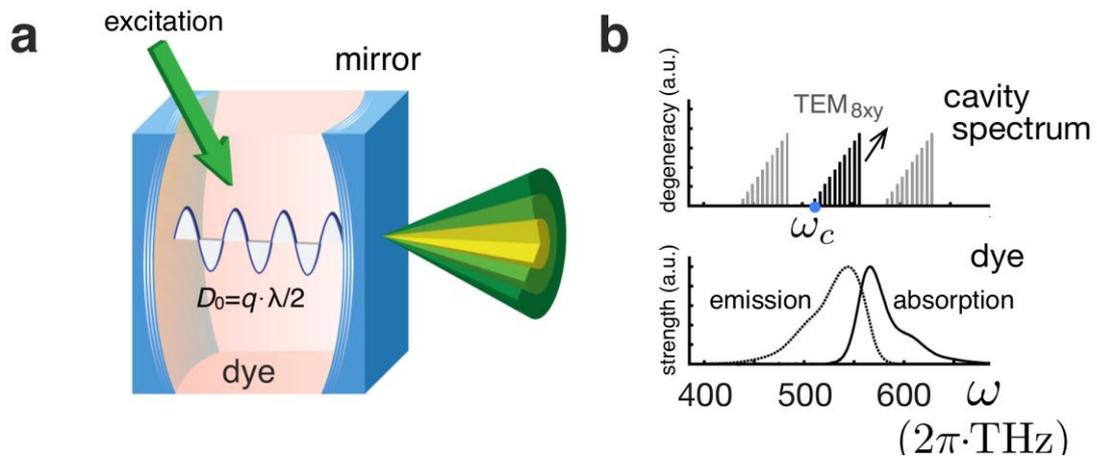

**Figure 1 | Bose-Einstein condensation of a two-dimensional photon gas, a,** Photons are captured inside a microcavity consisting of two spherically curved mirrors and get repeatedly absorbed and re-emitted by the embedded dye medium, leading to a thermalization of the photon gas to the temperature of the resonator (room temperature). **b**, The short cavity length causes a large frequency gap between the longitudinal resonator modes (free spectral range) of order of the emission bandwidth of the dye molecules. In this situation, the resonator becomes populated by photons of a single longitudinal mode number only, here $q$=8. However, the photons may still populate a multitude of transversally excited cavity modes (TEM$_{8xy}$ sub-spectrum), which effectively makes the photon gas two-dimensional.

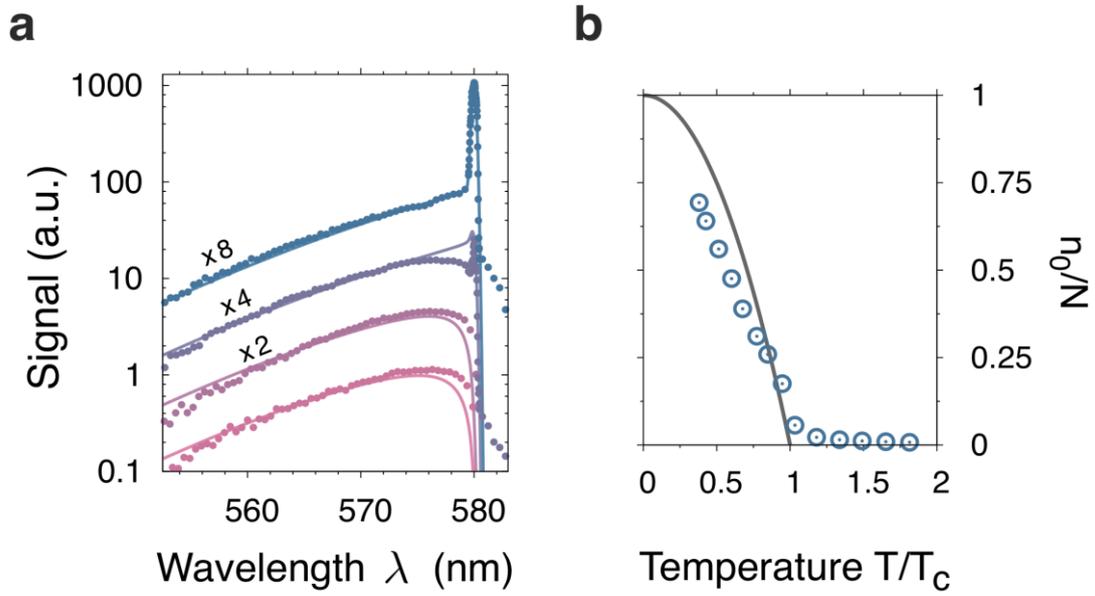

**Figure 2 | Spectral photon distribution and condensate fraction, a,** Distribution of photon energies for increasing total photon number (circles). For clarity the spectra have been vertically shifted. The observed spectra agree well with the expected 300K Bose-Einstein distribution functions (solid lines). **b,** Condensate fraction $n_0/N$ versus the reduced temperature $T/T_c$ along with the theoretical expectation $n_0/N=1-(T/T_c)^2$.

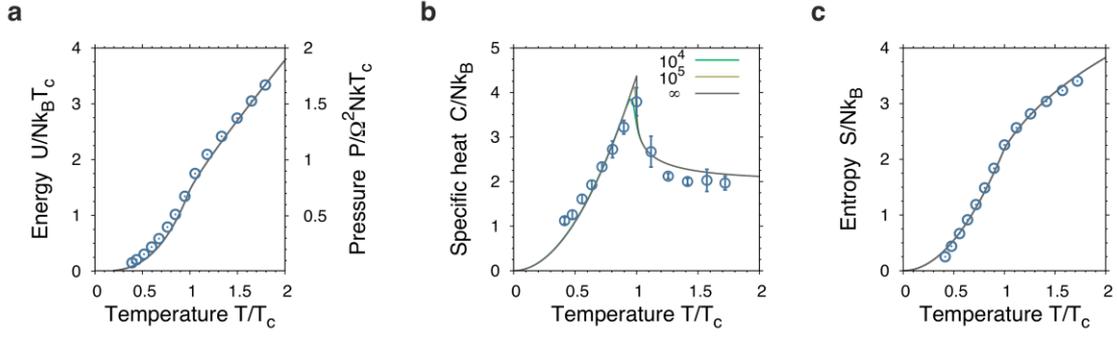

**Figure 3 | Internal energy, specific heat and entropy. a**, Internal energy (per photon) $U$ normalized to the characteristic energy $k_B T_c$ as a function of the reduced temperature $T/T_c$. In the classical high temperature regime, energy scales linearly with temperature as is expected from Maxwell-Boltzmann-like statistics. The formation of the condensate is signaled by a change in slope close to $T=T_c$. For an ideal two-dimensional trapped Bose gas, which is realized in our experiment in good approximation, the internal energy of the photon gas can moreover be linked to its pressure $P$ by $P=(1/2)\Omega^2 U$, see axis on the right hand side and the main text for details ($\Omega$ is the trapping frequency). Experiments are carried out at constant temperature $T=300$K and photon numbers ranging from N≈30,000 to $N$≈550,000, whereby an increase of the photon number corresponds to a decrease in the critical temperature $T_c=T_c(N)$ of the system. Statistical errors are smaller than the point size. **b**, Specific heat (per photon) versus the reduced temperature $T/T_c$, showing a cusp singularity at criticality (circles). In the high temperature (classical) regime, the specific heat reaches a limiting value of $2k_B$ per photon. At criticality $T=T_c$, the heat capacity reaches a maximum value of $C(T_c)/N=(3.8\pm0.3)$ $k_B$. The solid lines gives the specific heat of the two-dimensional harmonically trapped ideal Bose gas for varying total particle numbers. The error bars indicate statistical errors. **c**, Entropy per photon as a function of the reduced temperature $T/T_c$ (circles), along with the theoretically expected ideal Bose gas behaviour (solid line). The data is derived from a numerical integration of the specific heat curve of Fig. 3b (see text for details). The entropy monotonically decreases for decreasing temperatures, reaching a minimum value of ≈0.2$k_B$ per photon at the lowest achieved temperature. Statistical errors are smaller than the point size.